# Magnetotransport properties of $Cd_3As_2$ nanostructures

*Enze Zhang*[1,2†], *Yanwen Liu*[1,2†], *Weiyi Wang*[1,2], *Cheng Zhang*[1,2], *Peng Zhou*[3], *Zhi-Gang Chen*[4*], *Jin Zou*[4,5*] *and Faxian Xiu*[1,2*]

[1]State Key Laboratory of Surface Physics and Department of Physics, Fudan University, Shanghai 200433, China

[2]Collaborative Innovation Center of Advanced Microstructures, Fudan University, Shanghai 200433, China

[3]State Key Laboratory of ASIC and System, Department of Microelectronics, Fudan University, Shanghai 200433, China

[4]Materials Engineering, The University of Queensland, Brisbane QLD 4072, Australia

[5]Centre for Microscopy and Microanalysis, The University of Queensland, Brisbane QLD 4072, Australia

[*]Correspondence and requests for materials should be addressed to F. X. (E-mail: faxian@fudan.edu.cn), J. Z. (Email: j.zou@uq.edu.au) and Z-G. C. (Email: z.chen1@uq.edu.cn)

[†] These authors contributes equally to this work.

**ABSTRACT: Three-dimensional (3D) topological Dirac semimetal is a new kind of material that has a linear energy dispersion in 3D momentum space and can be viewed as an analog of graphene. Extensive efforts have been devoted to the understanding of bulk materials, but yet it remains a challenge to explore the intriguing physics in low-dimensional Dirac semimetals. Here, we report on the synthesis of $Cd_3As_2$ nanowires and nanobelts and a systematic investigation of their magnetotransport properties. Temperature-dependent ambipolar behavior is evidently demonstrated, suggesting the presence of finite-size of bandgap in nanowires. $Cd_3As_2$ nanobelts, however, exhibit metallic characteristics with a high carrier mobility exceeding 32,000 $cm^2V^{-1}s^{-1}$ and pronounced anomalous double-period Shubnikov-de Haas (SdH) oscillations. Unlike the bulk counterpart, the $Cd_3As_2$ nanobelts reveal the possibility of unusual change of the Fermi sphere owing to the suppression of the dimensionality. More importantly, their SdH oscillations can be effectively tuned by the gate voltage. The successful synthesis of $Cd_3As_2$ nanostructures and their rich physics open up exciting nanoelectronic applications of 3D Dirac semimetals.**

Dirac materials like graphene and topological insulators have been widely studied in recent years owing to their exciting physical properties originated from two-dimensional (2D) Dirac fermions.[1-5] In contrast with conventional semiconductors, their band structures obey a linear energy dispersion relation and possess vanishing effective mass near the Dirac point, thus promising applications in optoelectronics[6-9] and spintronics.[10-12] Recently, theory predicts the existence of three-dimensional (3D) Dirac fermions where the Dirac nodes are developed via the point contact of conduction-valence bands. The potential candidates involve β-$BiO_2$, $Na_3Bi$ and $Cd_3As_2$.[13-15] Interestingly, these 3D topological Dirac semimetals can be driven into topological insulators[14] or Weyl semimetals[14, 16, 17] by breaking symmetries that may lead to the discovery of novel physical phenomena such as quantum spin Hall effect and topological superconductivity.[14] Soon after the theoretical predictions, experiments like angle-resolved photoemission spectroscopy (ARPES)[3, 18-20] and scanning tunneling microscopy [21] were carried out on $Na_3Bi$ and $Cd_3As_2$ to investigate the 3D Dirac fermions.

$Cd_3As_2$ has been paid special attentions due to its chemical stability in air and extremely high mobility at both low and room temperatures.[20, 22] Studies based on $Cd_3As_2$ bulk materials have shown ultrahigh mobility of $9\times10^6$ $cm^2V^{-1}s^{-1}$ at 5 K[23] and up to $1.5\times10^4$ $cm^2V^{-1}s^{-1}$ at room temperature.[22] Giant and linear magnetoresistance[23, 24] and nontrivial π Berry's phase[25] of Dirac fermions were demonstrated in transport experiments. Recently, a superconductivity phase was also identified in a $Cd_3As_2$ crystal making it an interesting candidate of the topological superconductors.[26-29] However, despite the extensive studies on $Cd_3As_2$ bulk materials, few efforts have been devoted to the low-dimensional nanostructures, such as nanowires[30, 31] or nanobelts.

In this letter, we present the magnetotransport properties of $Cd_3As_2$ nanostructures. The superb crystallinity of as-grown $Cd_3As_2$ nanowires allows us to observe the semiconductor-like ambipolar effect, indicating the band gap opening. In contrast, $Cd_3As_2$ nanobelts exhibit high carrier mobility and pronounced double-period Shubnikov-de Haas (SdH) oscillations that can be effectively modulated by applied back-gate voltage ($V_{BG}$). The reduced dimensionality in these nanostructures, compared with the bulk crystals, offers the artificial suppression of Fermi sphere thus allowing for the observation of unique magnetotransport properties.

Transmission electron microscopy (TEM) was carried out to determine the structural characteristics of the synthesized $Cd_3As_2$ nanostructures. Figure 1a is a TEM image of a typical $Cd_3As_2$ nanowire with a diameter of ~80 nm. Figure 1b is a selected area electron diffraction (SEAD) pattern. The combination of Figure 1a and 1b indicates that the $Cd_3As_2$ nanowire is grown along [112] direction. The corresponding high-resolution (HR) TEM is depicted in Figure 1c, which clearly shows the $d_{112}$ spacing of 7.3 Å. Figure 1d displays another type of the $Cd_3As_2$ nanostructures − nanobelts − with an axial direction along [110], confirmed by the SAED pattern (Figure 1e), from which the large smooth surface of the nanobelt can be indexed as (112). The corresponding high-resolution (HR) TEM is shown in Figure 1f, which clearly shows the $d_{110}$ spacing of 4.6 Å.

To fabricate a field effect transistor (FET), the $Cd_3As_2$ nanowires were transferred onto a pre-patterned $SiO_2$/Si substrate. Optical microscopy and scanning electron microscopy (SEM) were subsequently used to locate the nanowires and determine their diameters. Source/Drain contacts were then deposited via a standard electron-beam lithography (EBL) process. Figure 2a inset shows the schematic drawing of the two-terminal FET device with a 40 nm-diameter $Cd_3As_2$ nanowire on top of a 285 nm $SiO_2$ gate (also see the SEM image, inset of Figure 2b). A $V_{BG}$ was applied on the degenerated silicon substrate to modulate the channel conductance. The gate-tuned source-drain current ($I_{DS}$) varies linearly with source-drain voltage ($V_{DS}$), indicating an Ohmic behavior (Figure 2a). A typical temperature-dependent resistance of the nanowires shows semiconducting characteristics (Figure 2b), indicative of a band gap opening[30, 32] which is reminisce of that in graphene nanoribbons.[33, 34] The activation energy can be further acquired by fitting the high-temperature resistance to equation $R_{xx}$ ~ $exp(E_a/k_BT)$, where $E_a$ is the activation energy and $k_B$ is the Boltzmann constant (Figure 2b inset). Here, $E_a$ is extracted to be 30 meV and the band gap $E_{gap}$ of the nanowire can be roughly estimated to be over 60 meV. To examine the temperature-dependent switching behavior of the $Cd_3As_2$ nanowire FET, the transfer curves of the device were obtained by sweeping $V_{BG}$ under a certain $V_{DS}$ of 5 mV from 2 to 120 K (Figure 2c). A strong ambipolar behavior is observed. The dashed guiding line in Figure 2c shows the shift of the threshold voltage $V_{TH}$ towards the negative direction as temperature increases. This is originated from the increasing electron charge carriers in the channel at elevated temperatures.[35] Also, temperature-dependent electron and hole mobility can be extracted from the linear region of the transfer characteristics using the equation,[36-38]

$$\mu_{FE} = [dI_{DS}/dV_{BG}] \times [L^2/(C_{ox}\, V_{DS})], \qquad (1)$$

where $\mathrm{d}I_{DS}/\mathrm{d}V_{BG}$ is the slope of the transfer curve in the linear regime, $L$ is the channel length, and $C_{ox} = (2\pi\varepsilon_0 \varepsilon L)/[cosh^{-1}(r + t_{ox}/r)]$ is the capacitance between the channel and the back gate through the $SiO_2$ gate dielectric ($\varepsilon_r$ = 3.9; $t_{ox}$=

285 nm; r is the radius of the $Cd_3As_2$ nanowire). As shown in Figure 2d, the mobility reaches the maximum value at ~20 K and decreases as temperate is increased which is attributed to the electron-phonon scattering.[39] The temperature dependence of mobility typically follows the relation $\mu \propto T^{-\gamma}$, where the temperature damping factor γ depends on electron-phonon coupling in the nanowire channel. By performing the best fit to the linear part of the curves (dash lines in Figure 2d), γ can be obtained to be ~0.5 and 1.7 for electrons and holes, respectively. Further suppression of phonon scattering can be realized by encapsulation of the nanowire in high-κ dielectric environment.[40]

Nanobelt is another form of nanostructures to study the physics in $Cd_3As_2$. To investigate the magnetotransport properties of the $Cd_3As_2$ nanobelts, back-gate devices with a standard Hall-bar geometry were fabricated, as schematically illustrated in Figure 3a (also see the SEM image, inset of Figure 3b). The longitudinal resistance ($R_{xx}$) as a function of temperature ($R_{xx}$-$T$) at zero magnetic field is acquired from the device with a 120 nm-thick $Cd_3As_2$ nanobelt (device 01, Figure 3b). Clearly, it shows a metallic behavior that is different from amorphous semiconducting $Cd_3As_2$ thin films.[41] In contrast with the quantum confinement effect in nanowires (gap opening), the $Cd_3As_2$ nanobelts are thick enough to incur the widely-observed band inversion (gapless),[14] leading to the high mobility and metallic characteristics. Figure 3c depicts the electron Hall slopes at different temperatures, from which the temperature-dependent Hall mobility and carrier density can be extracted (Figure 3d). The mobility reaches up to 32,000 $cm^2V^{-1}s^{-1}$ and drops as temperature rises because of the enhanced electron-phonon scattering.[40] We also observed a clear Hall anomaly and attribute it to the crystal anisotropy at low temperatures which is consistent with former transport experiments of $Cd_3As_2$ bulk materials.[23]

Owing to the high mobility of electrons in $Cd_3As_2$ nanobelts, we were able to resolve the ShH oscillations in the longitudinal magnetoresistance ($R_{xx}$) of the samples. Figure 4a displays the vertically-shifted $R_{xx}$ as a function of magnetic field $B$ at different temperatures. Surprisingly, pronounced double-period SdH oscillations persist up to 40 K. By subtracting the background, the plots of oscillation amplitude $\Delta R_{xx}$ versus $1/B$ from 2 to 40 K can be obtained, as depicted in Figure 4b. With the increase of temperature, the oscillation amplitude drops rapidly.[25] It is noted that the amplitude of the oscillations has a sudden increase when the magnetic field decreases to 4.5 T, suggesting two oscillation frequencies (Figure 4b). To obtain the oscillation frequencies $F$, we plot the Landau fan diagram by taking the maximum and minimum of the oscillation amplitude $\Delta R_{xx}$ as the half integer and integer, respectively (Figure 4c).[42, 43] The slopes yield two distinct frequencies $F_1$=66.8 T and $F_2$=63.8 T. By using the equation $F= (\phi_0/2\pi^2)S_F$, where $\phi_0=h/2e$, two cross-section areas of the Fermi surface $S_{F1}$= 6.40×$10^{-3}$Å$^{-2}$ and $S_{F2}$= 6.07×$10^{-3}$Å$^{-2}$ can be obtained. During the experiments, we also confirm that the magnetic field $B$ is perpendicular to [112] crystal direction that causes the development of two nested ellipses in the cross-section of Fermi surface, leading to the double-period anomalous SdH oscillations.[44-46] This is in a good agreement with previous ARPES results of two ellipsoidal Fermi surfaces.[18]

The observation of SdH oscillations is vital to analyze important parameters of the carrier transport.[47] In order to calculate the cyclotron mass ($m_{cyc}$), the temperature-dependent SdH oscillation amplitude $\Delta R_{xx}$ (Figure 4e) are extracted under zero back-gate voltage by fitting to the equation $\Delta R_{xx}(T)/R_{xx}(0)=\lambda(T)/sinh(\lambda(T))$, where the thermal factor is given by $\lambda(T)=2\pi^2 k_B T m_{cyc}/(\hbar eB)$, $k_B$ is the Boltzmann's constant, $\hbar$ is the reduced plank constant and $m_{cyc}=E_F/v_F^2$ is the effective cyclotron mass. We can obtain the effective cyclotron mass $m_{cyc}=0.046m_e$. Using the equation $v_F=\hbar k_F/m_{cyc}$, we can acquire the Fermi velocity $v_F$=1.15×$10^5$m/s and the Fermi energy $E_F$=342 meV. Moreover, from the Dingle plot the transport life $\tau$ time can be calculated from the Dingle factor exp(-D), where D=$2\pi^2 E_F c/(\tau e B v_F^2)$. Since $\Delta R_{xx}(T)/R$ is propositional to exp(-D)$\lambda(T)/sinh(\lambda(T))$, we can find $\tau$ from the slope of the logarithmic plot of $\Delta R_{xx}(T)/Rsinh(\lambda(T))$ versus $1/B$. By using extracted cyclotron masses the transport life $\tau$ can be estimated to be 1.59 ×$10^{-13}$s.[4, 48] Other important parameters such as mean free path $l=v_F\tau$ and the cyclotron mobility $\mu_{SdH}=e\tau/m_{cyc}$ are calculated, $l$=182 nm, $\mu_{SdH}$ =6144 $cm^2V^{-1}s^{-1}$, as summarized in Table 1.

To further probe the nature of SdH oscillations, the angular-dependent magnetotransport measurements were performed by tilting the sample from θ=0 °to 90 °. Figure 5a depicts $R_{xx}$ as a function of $B$ at varying tilting angle $\theta$, from which the SdH oscillations are observed to vanish when $\theta$ exceeds 45 °. After subtracting the background, the oscillation amplitude as a function of $B$ and $Bcos(\theta)$ was plotted in Figure 5b and c, respectively. It is known that for bulk $Cd_3As_2$ materials the SdH oscillations are observable from 0 °to 90 °owing to the nearly spherical Fermi surface.[23] While for the 2D electron gas in graphene or surface states in topological insulators the SdH oscillations show unchanged peak positions in the $Bcos(\theta)$ plot[48, 49]. Our results, however, are distinctive to both scenarios. Considering the thickness of 120 nm of the nanobelt, the disappearance of SdH oscillations at θ>45 °is attributed to the compressed Fermi sphere.[13]

To examine the tunability of back-gate voltage to the SdH oscillations, magnetotransport measurements were also carried out on 80 nm-thick $Cd_3As_2$ nanobelts (device 02). Figure 6a illustrates $R_{xx}$ as a function of back-gate voltage at 9 T. $R_{xx}$ decreases with the increase of $V_{BG}$, showing a clear *n*-type behavior. Interestingly, the resistance shows an oscillation-like feature at the high $V_{BG}$ regime, similar to that of 2D electron gas in black phosphorus.[50] Figure 6c illustrates the extracted SdH oscillations under different $V_{BG}$ at 2 K. Under zero or negative gate bias, the magnetoresistance does not show well-resolved SdH oscillations. While by applying a positive $V_{BG}$ of 20~60 V, pronounced SdH oscillations are obtained, suggesting the shifting of the Fermi level by back-gate voltage.[4, 51, 52]As showed in Figure 6b, the SdH oscillations' frequency increases from 42 to 70.1 T as $V_{BG}$ varies from 20 to 60 V. The calculated Fermi energy increases from 187 meV to 331 meV, revealing the lifting of Fermi level further into the conduction band.[52] The angular-dependent SdH oscillations under 60 V were then examined and there are no observable SdH oscillations as θ goes beyond 45 ° (Figure 6d). This is consistent with the device measured under zero back-gate voltage. To further understand the SdH oscillations, the Berry's phase can be obtained from the Landau fan diagram (Figure 6b). According to the Lifshitz-Onsager quantization rule,[43]

$$A_F \frac{\hbar}{eB} = 2\pi\left(n + \frac{1}{2} - \frac{\Phi_B}{2\pi}\right) = 2\pi(n + 1/2 + \beta + \delta), \qquad (2)$$

$2\pi\beta$ is Berry's phase, $2\pi\delta$ is the additional phase shift that changes from 0 for a quasi-

2D cylindrical Fermi surface to ±1/8 for a corrugated 3D Fermi surface.[25, 43] In our $Cd_3As_2$ nanobelt samples, under different $V_{BG}$ the intercept remains ~ 0.3, deviating from ±1/8, indicating a trivial zero Berry's phase. The presence of zero Berry's phase infers that the SdH oscillations mainly come from the high mobility conduction band and the Fermi surface is an anisotropic ellipsoid instead of sphere with prefect π Berry's phase.[44] It should be noted that with dimensionality reduced from bulk (3D) to nanowires (quasi-1D), $Cd_3As_2$ exhibits a transition from topological Dirac semimetal to trivial band insulator,[14] which makes the Dirac fermions disappear due to the band gap opening. At this stage, the detailed band structure and exquisite physics of the $Cd_3As_2$ nanostructures deserve further theoretical investigations.

In summary, we have incorporated the 3D Dirac semimetal $Cd_3As_2$ nanostructures into field-effect devices including nanowire FET and nanobelt back-gated Hall bar devices. The FET device exhibits ambipolar behavior, reminiscence of narrow-band-gap semiconductors. The gated-Hall bar devices, however, show ultrahigh mobility of $Cd_3As_2$ nanobelts and pronounced double-period SdH oscillations with reasonable gate tenability. Our results shed the light on the practical and versatile device applications of 3D Dirac semimetal in electronics.

## FIGURES

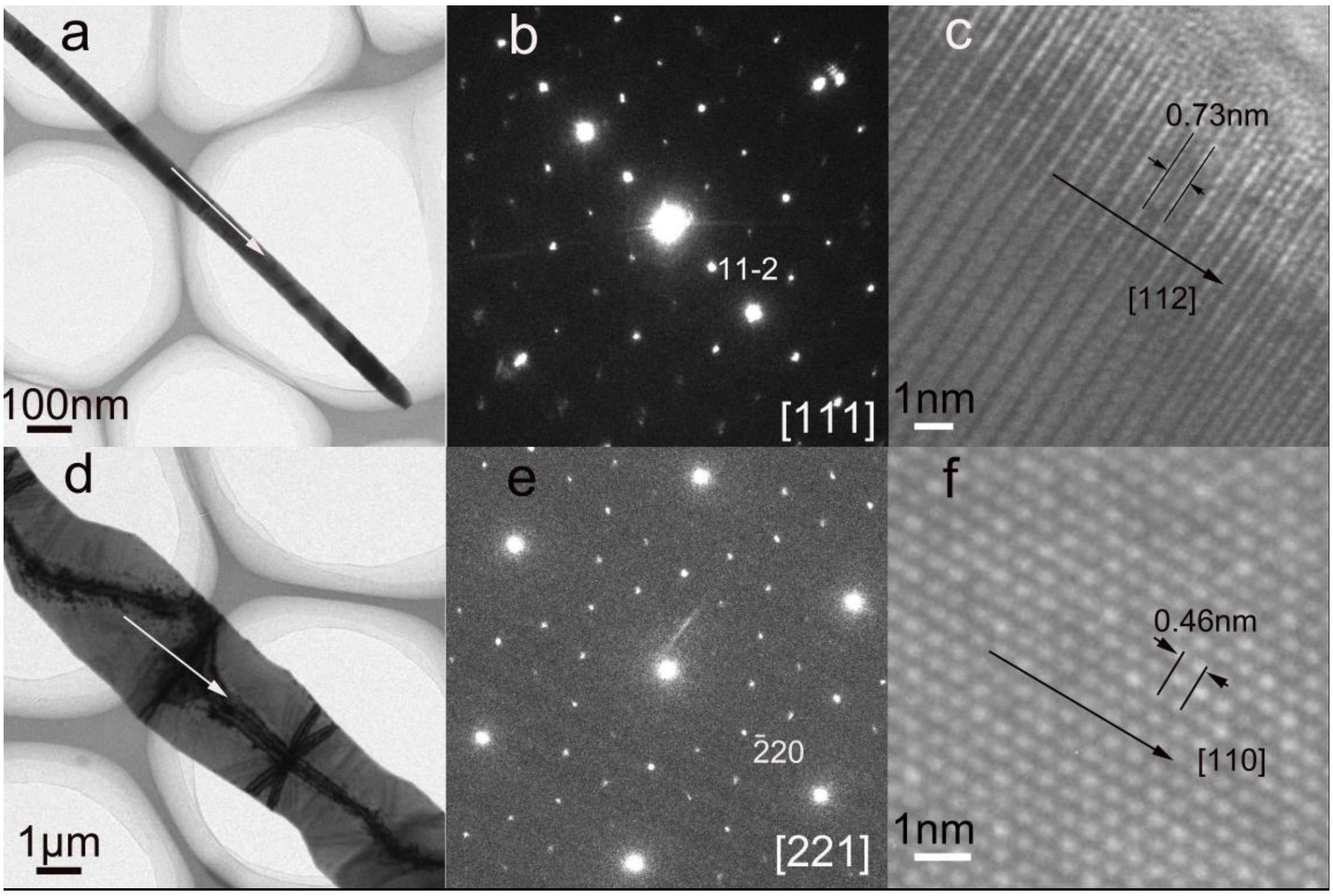

**Figure 1.** Microstructural characteristics of $Cd_3As_2$ nanostructures. (a) TEM, (b) SEAD, and (c) HRTEM images of the typical $Cd_3As_2$ nanowire; (d) TEM, (e) SEAD and (f) HRTEM images of the typical $Cd_3As_2$ nanobelt.

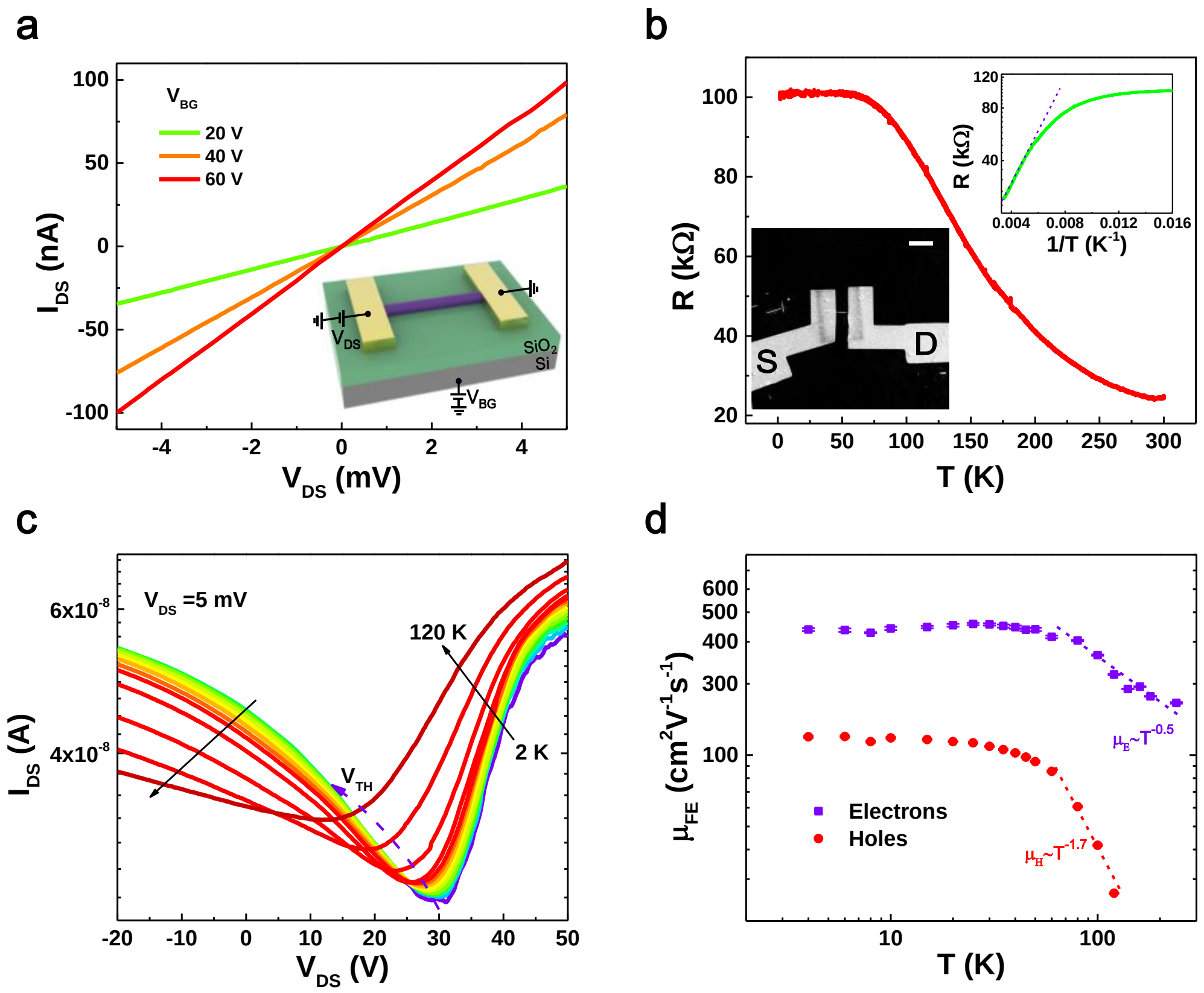

**Figure 2.** $Cd_3As_2$ nanowire field-effect transistor (FET) and its device characteristics. (a) The output characteristics ($I_{DS}$-$V_{DS}$) of the device under different back-gate voltages at 2 K. The inset shows a schematic structure of the fabricated device. (b) Typical channel resistance versus temperature (*R*-*T*) under zero back-gate voltage. The inset shows a SEM image of the FET device. Scale bar, 5 μm. (c) Temperature-dependent transfer curves of the device showing ambipolar effect at a fixed $V_{DS}$=5m V. (d) Field-effect mobility of electrons and holes as a function of temperature on a logarithmic scale.

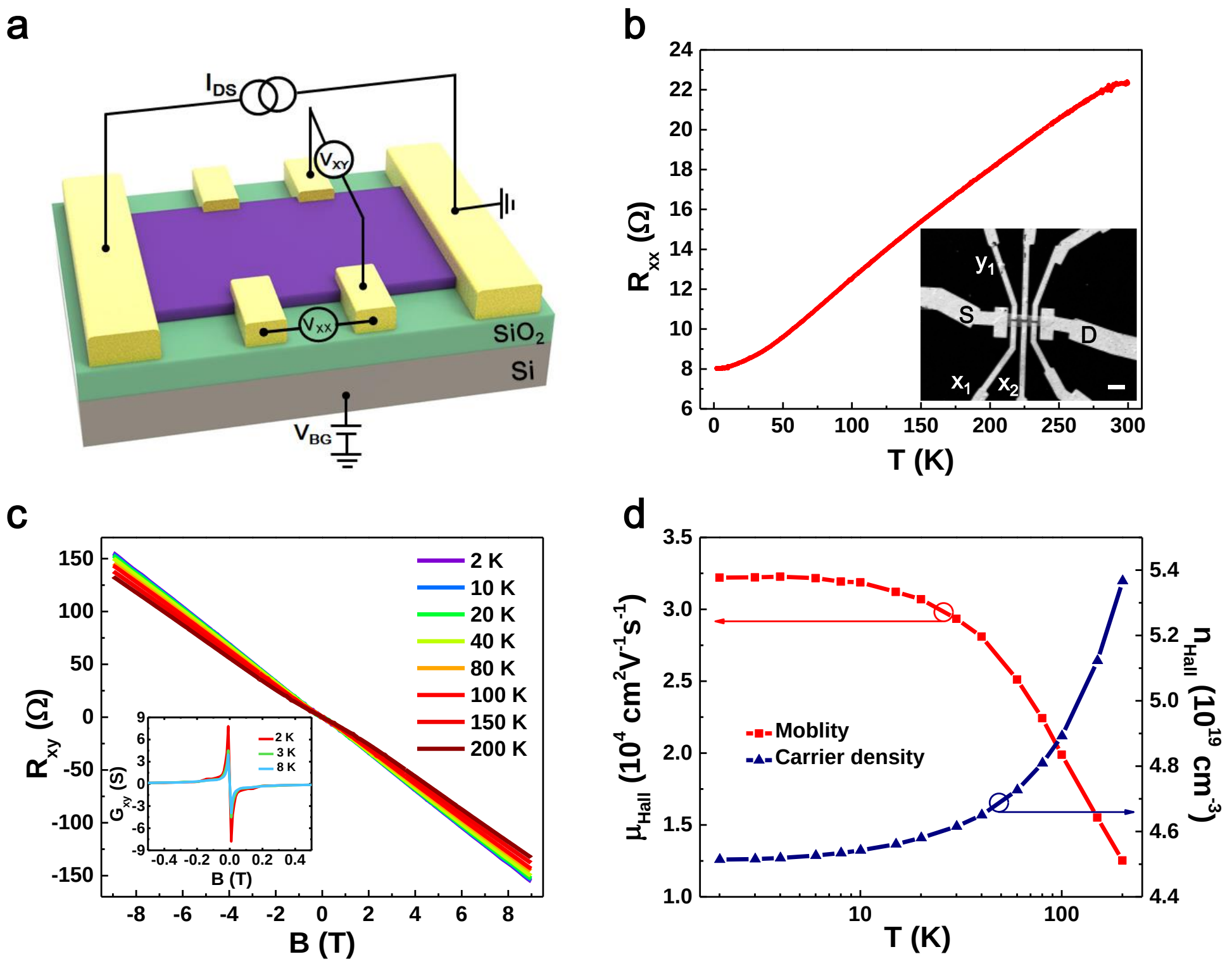

**Figure 3.** $Cd_3As_2$ nanobelt Hall-bar device and Hall effect characteristics. (a) Schematic device structure. (b) $R_{xx}$ as a function of temperature ($R_{xx}-T$) at zero magnetic field. The inset shows a SEM image of the device. Scale bar, 5 μm. (c) Hall resistance obtained at different temperatures. The inset shows a Hall anomaly of the device at low temperatures. (d) Hall mobility and carrier density deduced from the Hall slopes.

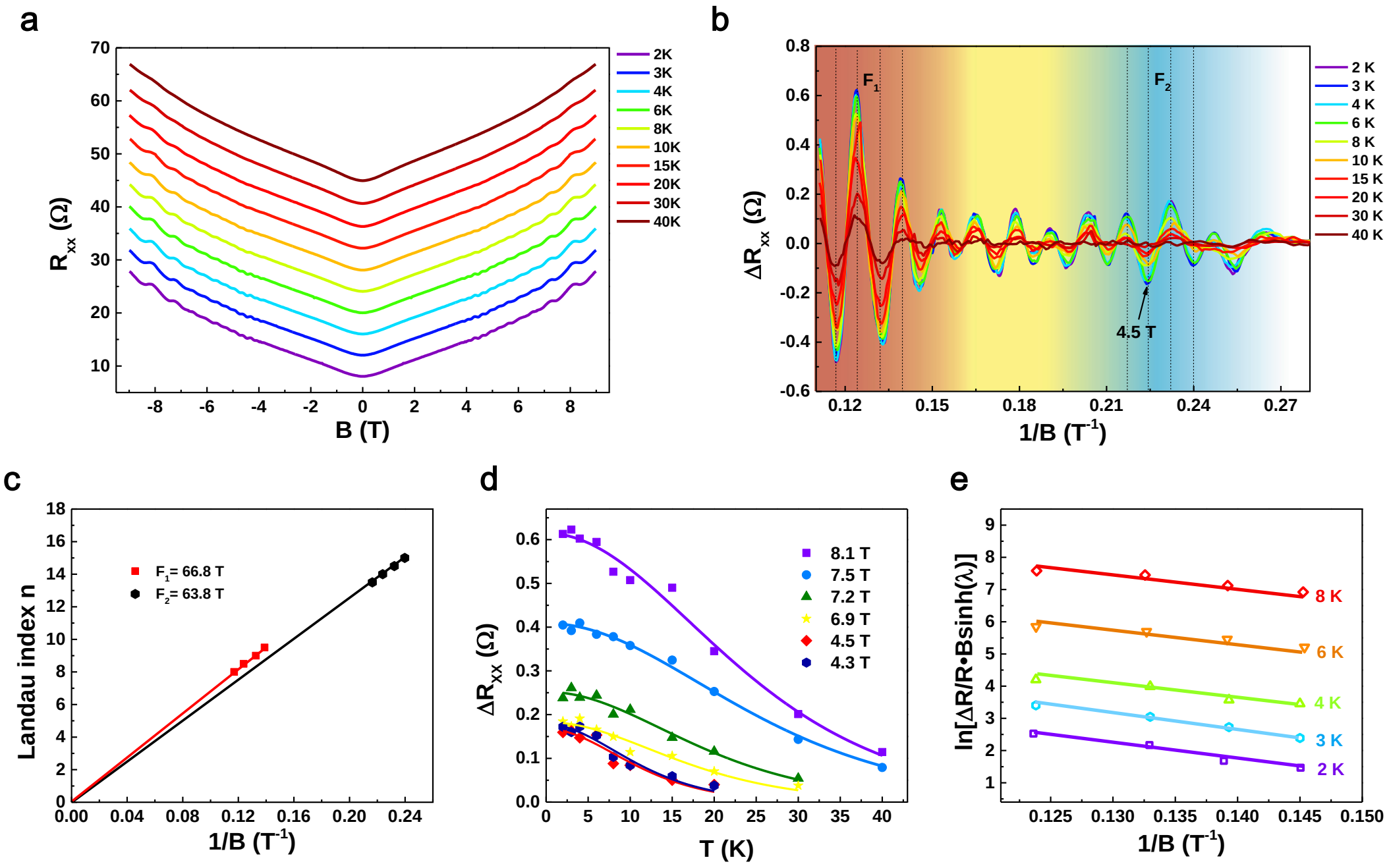

**Figure 4.** Temperature-dependent SdH oscillations of the $Cd_3As_2$ nanobelt Hall bar device. (a) Vertically-shifted $R_{xx}$ as a function of $B$ obtained at different temperatures showing two frequencies. (b) Temperature-dependent SdH oscillations' amplitude $\Delta R_{xx}$ as a function of $1/B$ after subtracting the background from a. (c) Landan fan diagram for SdH oscillations with two frequencies. The two intercepts are 0.28 and 0.31, which attribute to the trivial state. (d) $\Delta R_{xx}$ as a function of temperature. (e) Dingle plots for the extraction of transport life time.

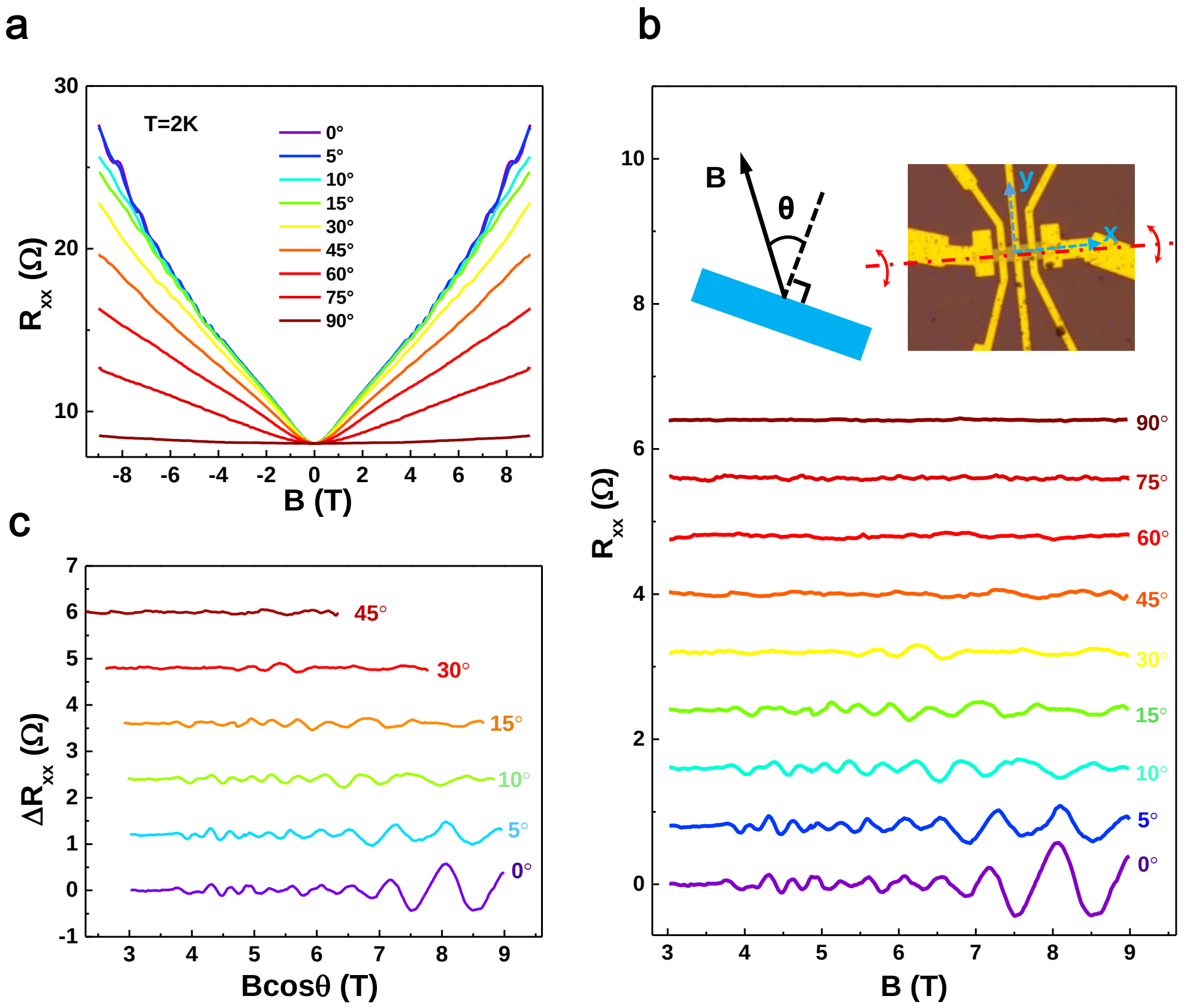

**Figure 5.** Angular-dependent SdH oscillations of the $Cd_3As_2$ nanobelt Hall bar device. (a) $R_{xx}$ as a function of $B$ at different tilt angle θ as defined in b inset. (b) SdH oscillations' amplitude $\Delta R_{xx}$ as function of $B$ after subtracting the background from a. Inset: measurement configuration. x and y indicate the crystal orientation of the $Cd_3As_2$ nanobelt. (c) SdH oscillations amplitude $\Delta R_{xx}$ as function of $Bcos(\theta)$

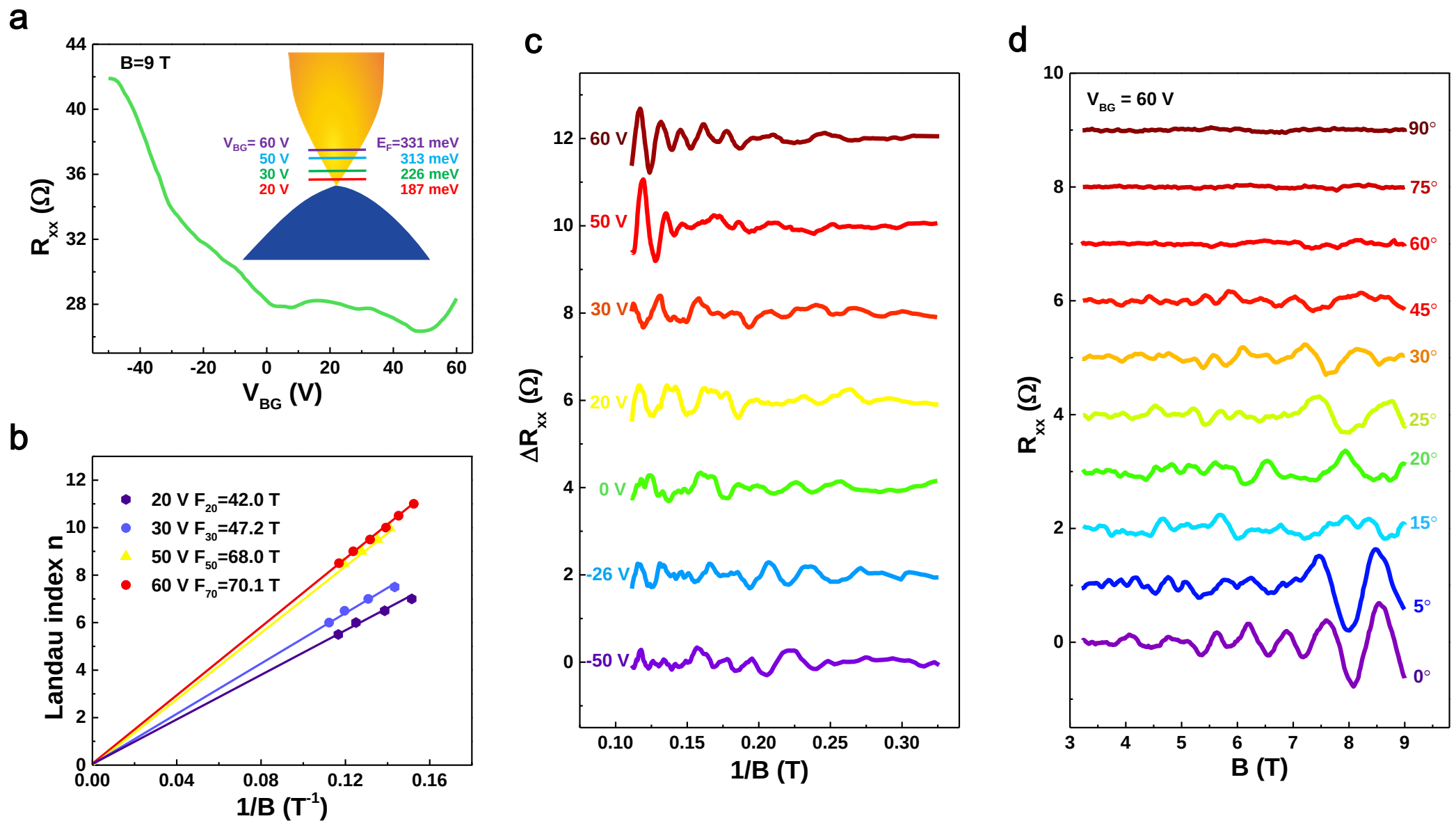

**Figure 6.** Gate-tuned SdH oscillations of the 80 nm-thick $Cd_3As_2$ nanobelt. (a) $R_{xx}$ as a function of $V_{BG}$ obtained at $B$=9 T. Inset: schematic band structure and the shifts of the Fermi level towards the conduction band at different $V_{BG}$. (c) Landan fan diagram of SdH oscillations for different $V_{BG}$. All the intercepts are around 0.3. (c) SdH oscillations amplitude $\Delta R_{xx}$ as function of $1/B$ obtained at different $V_{BG}$. (d) Angular-dependent SdH oscillations amplitude $\Delta R_{xx}$ as a function of $B$ at $V_{BG}$=60 V.

Table 1 Estimated parameters from the SdH oscillations

**Table 1 Estimated parameters from the SdH oscillations**

| Device label | $V_G$(V) | $F_{SdH}$ (T) | $S_f$ ($10^{-3}$Å$^{-2}$) | $K_f$ (Å) | $m_{cyc}$($m_e$) | $v_F$($10^6$ m/s) | $E_f$(meV) | t($10^{-13}$ s) | l(nm) | $\mu_{SdH}$(cm$^2$ V$^{-1}$s$^{-1}$) |
|---|---|---|---|---|---|---|---|---|---|---|
| 01 | 0 | 66.8 | 6.40 | 0.045 | 0.046 | 1.15 | 342 | 1.59 | 182 | 6144 |